\documentclass{iopjournal}

\usepackage{algpseudocode}
\usepackage{tikz}
\usetikzlibrary{shapes.geometric, positioning}
\usetikzlibrary{backgrounds}
\usetikzlibrary{calc}
\usepackage{amsmath}
\usepackage{amssymb}
\usepackage[numbers]{natbib}
\newcommand{\iotabar}{{\ooalign{\hidewidth\raisebox{0.1ex}{-}\hidewidth\cr$\iota$}}}

\usepackage{booktabs}          
\usepackage{amssymb}      
\usepackage[dvipsnames]{xcolor} 
\newcommand{\success}{{\color{ForestGreen}\checkmark}}
\newcommand{\failure}{{\color{BrickRed}\times}}

\begin{document}

\articletype{Paper} 

\title{Stellarator island divertor shape optimization for reduced peak heat fluxes}

\author{Avigdor Veksler$^1$, Aaron Bader$^2$, Heinke Frerichs$^3$, Elizabeth Paul$^1$}

\affil{$^1$Columbia University, New York NY, United States of America}

\affil{$^2$Type One Energy, Knoxville TN, United States of America}

\affil{$^3$University of Wisconsin --- Madison, Madison WI, United States of America}


\email{av3249@columbia.edu}

\keywords{fusion plasma, optimization, island divertor}

\begin{abstract}
An automated algorithm to construct island divertors for stellarators is presented and used to find divertors that meet heat load requirements determined by material limits. The algorithm uses just two initial conditions: two starting coordinates on the island separatrix chosen by the user. We leverage the simplicity of the algorithm to explore the divertor parameter space in a fixed magnetic equilibrium. Heat loads are approximated using the field line diffusion model implemented in the \texttt{FLARE} code. Divertor solutions that satisfy heat load requirements while maintaining a high power fraction captured are found using a parameter scan and a Bayesian optimization routine. The optimization finds divertors that perform the same as the parameter scan, but with a 75\% reduction in computational cost. The resulting divertors satisfy heat load requirements across varying cross-field heat diffusivities. Optimization over various islands in the equilibrium shows that it is easiest to find divertors that satisfy heat flux requirements for low-elongation islands. This algorithm presents a simple parameterization for island divertors and facilitates further physics and optimization studies.
\end{abstract}

\section{Introduction}
The fusion community has experienced an influx of fusion power plant (FPP) designs, driven by advances in computational techniques and record-high investment in private fusion companies \cite{fia_2024}. The stellarator community in particular has found advanced optimized equilibria \cite{landreman_2022, goodman_2023, ciemat_qi_2024}, making the stellarator a leading FPP candidate design for private companies. As more suitable stellarator equilibria are found, the  extrapolation of those equilibria to full-scale FPPs needs to be considered. This is typically where engineering and plant design meet the physics requirements of the FPP. One critical point of coordination between physics and engineering for stellarator FPPs is the power and particle exhaust solution. Fusion devices use a divertor to remove helium ash and other impurities. The divertor must also handle high heat fluxes from the incident thermal bulk plasma that is transported to the edge by drifts and collisional and turbulent processes. 

One divertor solution for stellarators is the island divertor, which was first demonstrated in W7-AS \cite{grigull_2001} and has been extensively studied in the W7-X experiment \cite{sunn_pedersen_first_2019, feng_understanding_2021, feng_modeling_2021, hammond_drift_2019}. The island divertor relies on stellarator equilibria with a resonance in the rotational transform profile at the edge of the plasma. This opens up a large edge island chain. Metal plates are placed to cut through the island poloidally. Plasma transported to the edge will enter the island chain and travel through it until it impacts the divertor plates. 

It is necessary for FPP divertors to be compatible with the intense heat fluxes coming from the incident plasma. Currently, materials used to construct divertors, such as tungsten, are rated to sustain steady state heat loads of $10~\text{MW}/\text{m}^2$ \cite{escourbiac_2019}. Exposure to heat loads greater than $10~\text{MW}/\text{m}^2$ may lead to cracking and recrystallization, which degrades the material's ability to withstand high thermal loads over time \cite{jin_influence_2022}. Additional heat sources on the divertor may come from energetic alpha particles that can deposit heat flux on the divertor plate \cite{liu_loss_2022}. Furthermore, there is a non-negligible heat flux resulting from neutrons in a burning plasma \cite{norajitra_divertor_2011}. Recent works \cite{you_high_heat_flux_2022, richou_development_2026} show that divertor designs can withstand higher heat fluxes, up to $20~\text{MW}/\text{m}^2$, but divertor component lifetime is inversely proportional to the incident heat flux, with lower heat fluxes corresponding to longer divertor lifetimes, as the plasma facing components and their cooling structures experience less fatigue  \cite{divertor_lifetime_2018}. Low heat fluxes also lead to reactor scenario flexibility. If the heat flux is reduced as much as possible for a given reactor scenario, the reactor radiative fraction requirements can be reduced, leading to better core performance and greater output power. Therefore, a divertor with lower heat flux is beneficial for reactor scenario design.

One benefit of stellarator island divertors is that the long connection lengths of stellarator islands help reduce the peak heat loads \cite{feng_comparison_2011}. For stellarators, the target-to-target connection length measures the distance a field line travels from one surface before terminating on a second surface, such as a divertor plate \cite{feng_review_2022}. These connection lengths determine the parallel transport time scales and hence the heat flux spread perpendicular to the magnetic field due to diffusive processes \cite{feng_review_2022}.

Stellarator FPP solutions with planned island divertors should be designed with long connection lengths in the island region, since long connection lengths promote more perpendicular diffusion and lower peak heat loads on the divertor plates. Another way to reduce peak heat loads is to target a detached divertor regime, where upstream neutrals dissipate plasma power before it reaches the plate \cite{Krasheninnikov_Kukushkin_2017}. The physics of detachment are not considered in this paper where we only consider an incident heat flux due to non-radiated, scrape-off-layer power, $P_\text{SOL}$, which may be achieved with some combination of core and edge radiation. Yet another way to reduce the peak heat loads is to shape the divertor plate with respect to the magnetic field geometry. Ensuring a small angle of incidence of the magnetic field to the divertor plate helps spread out the incident heat loads across the plate as much as possible. The heat flux limit is a mandatory requirement for a survivable divertor solution. The divertor's functionality is defined through particle exhaust and impurity retention characteristics. A high exhaust efficiency is a performance requirement that should be maximized. The first principle of this is to get particles onto the target plate to neutralize, which is achieved by enforcing a high fraction of power captured on the divertor plates as opposed to other surfaces. 

There is previous work on divertor shape optimization. For tokamaks, adjoint-based methods were applied to optimize the shape of target plates using reduced edge physics models \cite{dekeyser_divertor_2014, baelmans_achievements_2017}, and recently, population-based methods, using field line diffusion models to estimate heat loads, were used to optimize divertor geometries \cite{frerichs_particle_2025}. For the stellarator island divertor, previous research on W7-X explored the effect of magnetic geometry shaping with respect to the island divertor plate \cite{pedersen_first_2018}. There are various geometry generation algorithms for divertor plates that target divertor heat load minimization \cite{davies_semi-automated_2024, liu_2024}, and most recently neutral models have been implemented into optimization of the W7-X divertor \cite{frerichs_2026}. These represent important steps in divertor optimization.

The purpose of this work is to present a low-dimensional parametrization that can be used to find divertor geometries that spread out the heat flux by maintaining a low incident angle with respect to a fixed background magnetic field equilibrium. Details of the example equilibrium used can be found in \citet{hegna_2025}. The low dimensionality makes it suitable for parameter scans and optimization. Bayesian optimization is used as a proof-of-principle to optimize the divertor geometry to minimize the peak heat loads as much as possible while maintaining a high fraction of power captured by the divertor plates. New computational capabilities in the \texttt{FLARE} code \cite{frerichs_magnetic_2025} make it possible to use expected heat loads as part of an optimization loop. A neutral model is not included in this paper, but would be an important part of divertor optimization to ensure that the optimization solutions work with regard to particle exhaust and retention. Despite this, this work is valuable in outlining how a simple divertor geometry generator can be parametrized for and interfaced with an optimization algorithm, and is a step towards a comprehensive divertor design process that is necessary for full-device optimization \cite{gates_recent_2017}.

In Sec.~\ref{sec:Methods}, the algorithm to automatically generate divertors is described, as well as the numerical simulation and cost function used to evaluate divertor compatibility with respect to the heat flux and power fraction captured requirements. The divertors that result from the integration of the algorithm with a Bayesian optimization loop and their physics analysis are shown in Sec.~\ref{sec:results}. We conclude in Sec.~\ref{sec:conclusions}.

\section{Methods}\label{sec:Methods}

\begin{figure}
    \centering
    \includegraphics[width=0.7\linewidth]{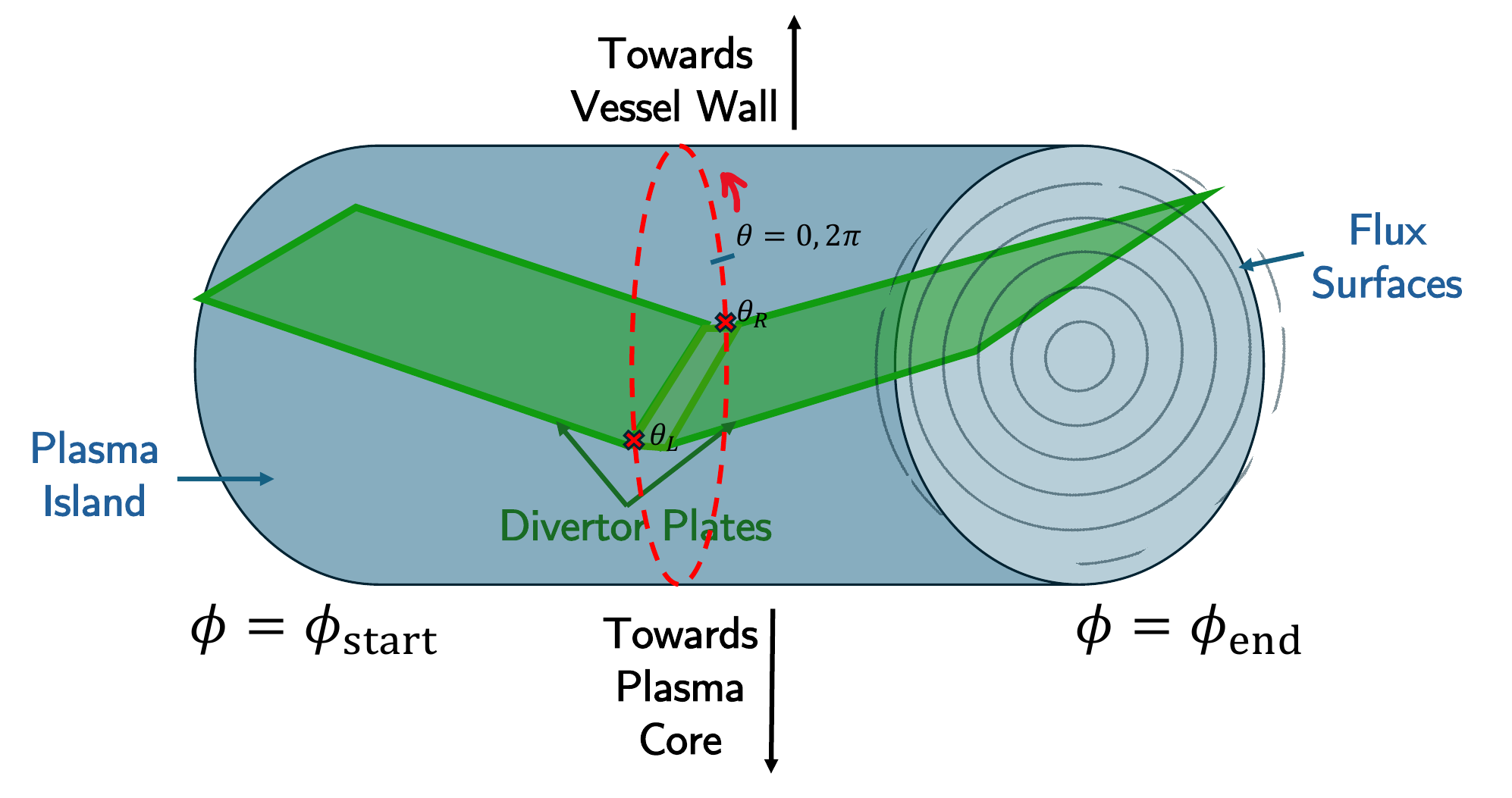}
    \caption{A cartoon of a V-shaped island divertor (green) in a stellarator island (blue cylinder, representative of a single stellarator island). Although in this cartoon the plates are shown to be planar, the ones created by the algorithm described in Fig.~\ref{fig:divertor_making_algorithm} allow for curved geometries in order to target the proper strike angle at both points of contact with the island separatrix. The divertor extends from $\phi=\phi_{\text{start}}$ to $\phi=\phi_{\text{end}}$, where $\phi$ is the toroidal angle of the stellarator in standard cylindrical coordinates. A rounded junction where the plates meet avoids having a sharp discontinuity in the divertor. The initial starting angles that define the divertor geometry are shown as well. The algorithm is described in further detail in Sec.~\ref{subsec:div_making_algo}.}
    \label{fig:cartoon_divertor}
\end{figure}

The divertor-making algorithm described below creates an island divertor made up of straight segments at constant $\phi$. The divertor geometry has curvature as these straight segments will vary in orientation as $\phi$ changes. In the toroidal direction, the divertor has a V-shape, where the tip of the V is closest to the core plasma. A cartoon image of this is shown in Fig.~\ref{fig:cartoon_divertor}. The rest of the divertor plates slope away from the core towards the confinement vessel wall. This orientation catches heat flux from particles that are co- and counter-passing with respect to the magnetic field. It was also found that this orientation tended towards fewer leading edges and larger divertor wetted areas. Although this introduces some divertor self-shadowing, the divertor plates from the computational design can be cut to only include the portions with expected heat flux. This, however, may not be ideal for a reactor where fluctuations in stellarator edge parameters during startup could deposit heat flux over a broader area, as detailed in Sec.~\ref{subsec:divertor_robustness}.
\subsection{Equilibrium parameters}
Detailed information for the equilibrium used for this study can be found in \citet{hegna_2025}. The equilibrium examined has 4 field periods, is stellarator symmetric, and has an edge iota of $\iotabar = 0.8$, yielding a $n=4$, $m=5$ resonant island chain for an island divertor. The equilibrium is scoped to produce $800~\text{MW}$ $P_\text{fusion}$, which yields $160~\text{MW}$ of plasma heating due to alpha particles \cite{bader_power_2025}. In this work, we use a radiated power fraction of $60\%$ to set $P_\text{SOL}=64~\text{MW}$, which is the non-radiated power that must be captured by all of the divertor units. This is a conservative estimate, but one that will highlight the algorithm's ability to design divertors that are suitable for FPP scale heat fluxes. Since the equilibrium is stellarator symmetric, we enforce that the divertor lies in a half-field period of the equilibrium magnetic field used in this work. This symmetry is exploited to reduce computational costs in Sec.~\ref{sec:results}. A single divertor is modeled in a half-field period, and $P_\text{SOL}/(2N_p)$, where $N_p$ is the number of field periods, is carried by fieldlines that diffuse from the last closed flux surface to the divertor in the half-field period.

\subsection{Divertor-making algorithm}\label{subsec:div_making_algo}

In this section, we present a simple, low-dimensional parametrization of divertor construction. This algorithm creates divertors that aim to minimize the peak heat loads on the divertor by enforcing a maximum strike angle of the magnetic field on the divertor plates near the island separatrix. Maintaining a low strike angle helps spread the heat flux across the plate by increasing the area that is exposed to the plasma. Although very small angles are a feature of current island divertors, as in W7-X, there are risks in attempting to build divertors with such shallow angles in terms of engineering tolerances \cite{wai_angle_limit_2023}. In principle, a minimum strike angle could be enforced with further adjustments of the divertor plate files.  

\begin{figure*}
    \centering
    \begin{tikzpicture}[
        node distance=0.8cm and 1cm,
        auto,
        block/.style={
            rectangle,
            draw,
            thick,
            text width=8.5cm,
            align=center,
            rounded corners,
            fill=gray!10,
            font=\sffamily
        },
        decision/.style={
            rectangle,
            draw,
            thick,
            text width=3cm,
            align=center,
            rounded corners,
            fill=gray!10,
            font=\sffamily
        },
        answer/.style={
            rectangle,
            draw,
            thick,
            minimum width=1.5cm,
            align=center,
            fill=gray!10,
            font=\sffamily\bfseries
        },
        line/.style={
            draw,
            ->,
            thick
        }
    ]
    
    
        \node [block] (init) {
            $\phi = \phi_{\text{init}}$\\ 
            $p_{L}=(\phi, \theta_{L0})$\\
            $p_{R}=(\phi, \theta_{R0})$
        };
    
        \node [block, below=of init] (trace) {
            Follow magnetic field from point $p_{L}$
            to the $\phi+\Delta\phi$ plane. Call this point $p_{L,b}$
        };
    
        \node [block, below=of trace] (find_left) {
            Search for a point $p_L^* =(\phi+\Delta\phi, \theta_L^*)$ on the $\phi+\Delta\phi$ island separatrix  near $p_{L,b}$ such that the magnetic field $\vec{b}$ evaluated halfway between $p_{L}$ and $p_L^*$ on the vector $\vec{d}_L=p_L^*-p_L$ satisfies $\hat{b}_L \cdot \hat{d}_L = \cos(\alpha)$
        };
    
        \node [block, below=of find_left] (find_right) {
            Complete same procedure for right point.
        };
    
        \node [block, below=of find_right] (update) {
            Update $p_L=p_L^*$, $p_R=p_R^*$, and save them to a list.
        };
    
        \node [decision, below=of update] (check_phi) {
            $\phi = \phi_{\text{stop}}?$
        };
    
        \node [answer, below=0.6cm of check_phi] (yes_box) {Yes};
        \node [answer, right=4cm of check_phi] (no_box) {No};

        \node [block, text width=2cm, above=2cm of no_box] (update_phi) {
            Update $\phi = \phi + \Delta\phi$
        };
    
        \node [block, below=0.6cm of yes_box] (compute_vec) {
            For all pairs of $p_L$ and $p_R$ points, compute additional points along line segment connecting $p_L$ and $p_R$ to extend the divertor plate past the width of the island.
        };
    
        \node [block, below=of compute_vec] (final_save) {
            Save those points in Kisslinger format.
        };
    
    
        \path [line] (init) -- (trace);
        \path [line] (trace) -- (find_left);
        \path [line] (find_left) -- (find_right);
        \path [line] (find_right) -- (update);
        \path [line] (update) -- (check_phi);
        \path [line] (check_phi) -- (yes_box);
        \path [line] (check_phi) -- (no_box);
        \path [line] (yes_box) -- (compute_vec);
        \path [line] (compute_vec) -- (final_save);
        
        \path [line] (no_box) -- (update_phi);
        \draw [line] (update_phi.north) |- (trace.east);
    \end{tikzpicture}
    \caption{Algorithm that creates a divertor plate geometry from $\phi_{\text{init}}$ to $\phi_{\text{end}}$. The algorithm is run with $\pm\Delta\phi$ to create a V-shaped divertor.}
    \label{fig:divertor_making_algorithm}
\end{figure*}

The divertor-making algorithm is described in Fig.~\ref{fig:divertor_making_algorithm}. The inputs required are: a discretization of points describing the island separatrix from $\phi_{\text{start}}$ to $\phi_\text{end}$ in increments of $\Delta\phi$, the capability to evaluate the plasma magnetic field beyond the LCFS of the core plasma, and two starting points on the island separatrix. In reality, a flux surface within the island near the island separatrix is used, called the control surface, as magnetic chaos could complicate the analysis if the actual separatrix were chosen. Only the control surface is used to construct the divertor. The rationale for this is that most of the heat flux will travel on flux tubes near the island separatrix towards the divertor plate \cite{Kharwandikar2026, effenberg_investigation_2019}. The number of points describing the control surface per toroidal angle should be enough to be able to build a smooth periodic spline representation of the surface.

The output of the algorithm is two divertor plate files in Kisslinger format \cite{feng_review_2022}, each representing a side of the V-shaped divertor. The files are text files that describe the geometry of divertor plates from $\phi\in[\phi_\text{start}, \phi_\text{init}]$ and from $\phi\in[\phi_\text{init}, \phi_\text{end}]$, where $\phi_\text{start}$ and $\phi_\text{end}$ are defined by the user. $\phi_\text{init}$ is the toroidal location with maximal radial insertion into the island. In this work, it was placed at the midpoint $\phi_\text{init}=(\phi_\text{start}+\phi_\text{end})/2$, but this does not have to be the case, and in general the plane at $\phi_\text{init}$ is not a symmetry plane.

\begin{figure}
    \centering
    \includegraphics[width=0.7\linewidth]{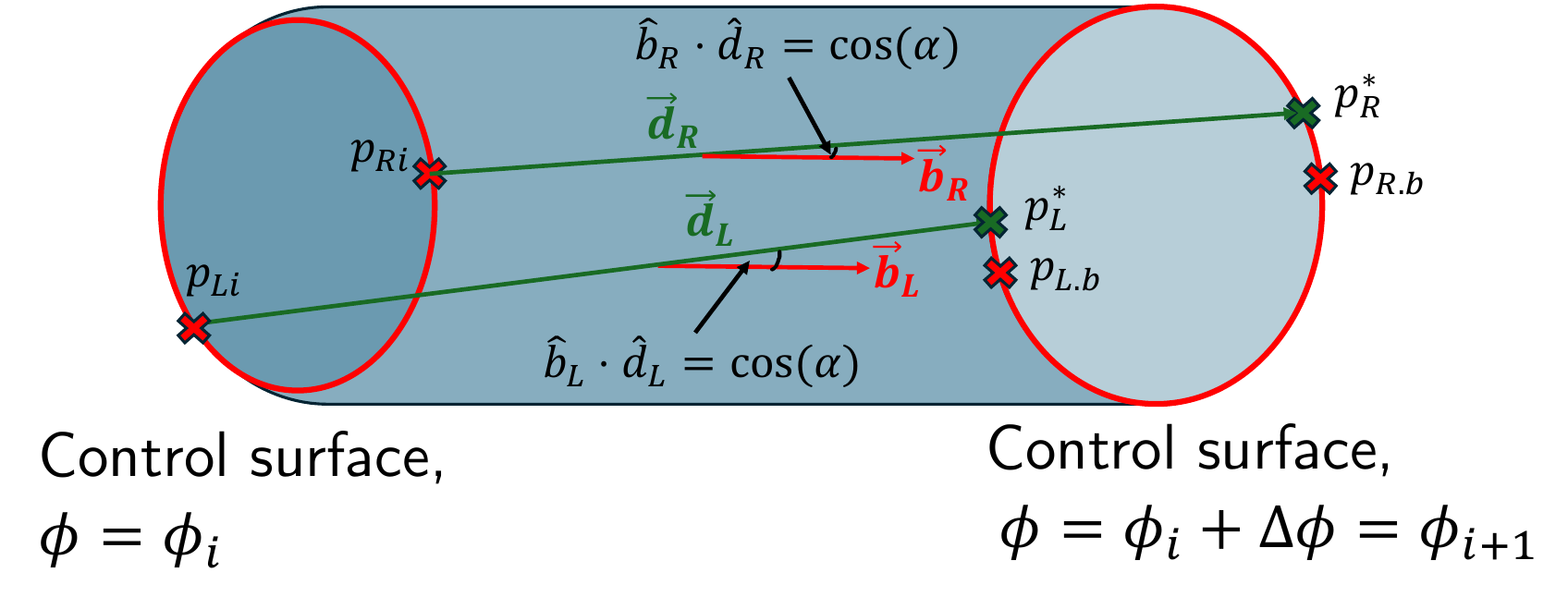}
    \caption{A visualization of how the coordinates of the divertor intersections with the control surface at different $\phi$ planes are chosen. The point $p_L^*$ is found near the point $p_{L.b}$ such that $\hat{b}_L\cdot \hat{d}_L=\cos(\alpha)$. The same is repeated for the right points, and the divertor section between $\phi_i$ and $\phi_{i+1}$ is a surface going through $p_{Li}, p_L^*, p_{Ri}, p_R^*$.}
    \label{fig:cartoon_divertor_algorithm}
\end{figure}

The algorithm is a low-dimensional parametrization of divertor construction as there are only two inputs, the initial $(p_{L0}, p_{R0})$ starting points on the $\phi_\text{init}$ control surface. Let the initial starting points on the $\phi_\text{init}$ control surface be $p_{Li}=p_{L0}=(r_{L0}, \phi_{\text{init}}, z_{L0})$ and $p_{Ri}=p_{R0}=(r_{R0}, \phi_{\text{init}}, z_{R0})$ in a standard cylindrical coordinate system. The index $i$ denotes the toroidal plane, starting from $i=0$. In practice, it is convenient to have a continuous representation of the initial conditions by mapping the starting input coordinates to angles using a periodic B-spline representation of the control surface at $\phi_\text{init}$. The spline is generated over a list of angles defined by 
\begin{equation*}
    \theta= \mathrm{arctan2}(z-z_0, r-r_0)
\end{equation*}
where $(r,z)$ are coordinates of the control surface and $(r_0, z_0)$ are the coordinates of the O-point of the island. Thus, a pair of angles $(\theta_L, \theta_R)\in[0, 2\pi)\times[0, 2\pi)$ on the $\phi_\text{init}$ plane represent starting coordinates $p_{L0}=(\phi_\text{init}, \theta_{L0})=(r_{L0}, \phi_\text{init},z_{L0}), p_{R0}=(\phi_\text{init}, \theta_{R0})=(r_{R0}, \phi_\text{init}, z_{R0})$. The angles $(\theta_L, \theta_R)$, which are defined per control surface, are now a continuous, two dimensional input parameter space for the algorithm.  From these two points, the rest of the divertor geometry is automatically generated.  We integrate the magnetic field trajectory from $p_{Li}$ and $p_{Ri}$ to the $\phi+\Delta\phi$ plane. These points also lie on the control surface, and we call them $p_{L,b}$ and $p_{R,b}$. Now we search for points near $p_{L,b}$ and $p_{R,b}$ called $p_L^*$ and $p_R^*$. We pick the points $p_j^*, j=L,R$ such that the magnetic field vector $\vec{b}_j$ evaluated at $\vec{r}_j=\frac{1}{2}(\vec{p}_{ji}+\vec{p}_j^*)$, the midpoint between $p_{ji}$ and $p_j^*$ along the vector $\vec{d}_j=\vec{p}_{ji}-\vec{p}_j^*$, satisfies $\hat{b_j}\cdot\hat{d_j}=\cos(\alpha)$, where $\alpha$ is the maximum strike angle specified for the divertor-making algorithm. A visualization of this process for the left point is shown in Fig.~\ref{fig:cartoon_divertor_algorithm}. The section of the divertor between $\phi$ and $\phi+\Delta\phi$ will be a surface going through $p_{Li}$, $p_L^*$, $p_{Ri}$, and $p_R^*$. This enforces that the magnetic field (and approximately, the incoming particle heat flux) would have a maximum strike angle of $\alpha$ with the divertor in the toroidal region $\phi$ to $\phi+\Delta\phi$.

The algorithm then steps toroidally in $\phi$. The same procedure above is used but now from the $\phi+\Delta\phi$ plane, using $p_{j,i+1}=p_j^*$. This is repeated until we reach $\phi=\phi_{\text{stop}}$. The list of points $[p_{j0}, \dots, p_{jN}]$, where $N=|(\phi_{\text{init}} - \phi_{\text{stop}})|/\Delta\phi$, is saved. The algorithm is run twice, once where $\Delta\phi$ is positive and the stopping criterion $\phi_{\text{stop}}=\phi_\text{end}$ and once where $\Delta\phi$ is negative and the stopping criterion is $\phi_\text{stop}=\phi_\text{start}$. For every $\phi$ plane, additional points are used to extend the divertor plate along the line segment connecting $p_{L}, p_{R}$ by $0.2*|p_{L}-p_{R}|$. This ensures that the divertor captures the heat flux spreading in the region outside of the island. 

At the end of the divertor-making algorithm, we create two additional divertor plate files. These are flat plates in the $(r, z)$ plane that ensure no particles hit behind the divertor, and are described in more detail in Sec.~\ref{subsec:divertor_performance}.

The strike angle $\alpha$ is also a user-chosen parameter. In this work the maximum strike angle was set to $3^\circ$ to spread the heat flux along the toroidal extent of the divertor plate under attached conditions. 

In order to avoid a sharp discontinuity at $\phi_\text{init}$ where the two plates meet, the algorithm uses an input maximum strike angle $\alpha$ that increases linearly from $0.1^\circ$ to $3^\circ$ over the first four $\phi$ planes before and after $\phi_\text{init}$. This creates a gradual slope where the two divertor plates meet.

\subsection{\texttt{FLARE} code}\label{sec:flare_code}

The \texttt{FLARE} \cite{frerichs_flare_2024} code is a magnetic meshing and field line analysis code, and is used to approximate heat flux loads on the divertors designed by our divertor-making algorithm. \texttt{FLARE} has an adaptive magnetic mesh generation algorithm \cite{frerichs_magnetic_2025}. The magnetic mesh consists of magnetic flux tubes which can quickly reconstruct field line positions using a local flux-tube coordinate system. The magnetic mesh is critical as it provides a significant speed up for field line tracing, around 100 - 1000 times faster \cite{frerichs_magnetic_2025} when compared to numerical integration of particle trajectories. 

The other components of \texttt{FLARE} used are the field line tracing modules. Field line tracing can be used for simple models of heat flux transport. \texttt{FLARE} employs a field line diffusion model, which models heat transport via parallel streaming of particles along fieldlines and perpendicular diffusive kicks to model cross-field transport. Field line diffusion models are Monte-Carlo methods for the source-free continuity equation \cite{feng_review_2022}, solving
\begin{equation}
    \vec{\nabla}\cdot(nC_s\vec{b}-D\vec{\nabla}_{\perp}n)=0
    \label{eqn:continuity}
\end{equation}
where $n$ is the ion density, $C_s$ is the ion sound speed, and $D$ is a diffusion coefficient. In simulation, after a given parallel integration length $\Delta l=C_s\tau$ with timestep $\tau$, the particles are kicked in the perpendicular direction by $\vec{\Delta}_{\perp}=\sqrt{4D\tau}\vec{\xi}_\perp$, where $\vec{\xi}_\perp$ is a randomly chosen unit vector perpendicular to the local magnetic field. Details on the implementation of the field line diffusion algorithm can be found in \citet{frerichs_magnetic_2025}. Proxies for heat loads on divertor plates can be estimated by initializing particles on the LCFS, and tracing their diffusive trajectories into the island and onto the divertor plates.

\texttt{FLARE}'s field line diffusion model can be initialized by specifying either a cross-field diffusion coefficient in terms of spatial steps, used in Sec.~\ref{subsec:divertor_robustness}, or a heat flux diffusivity $\chi_\perp$ that is used in a similar field line diffusion model in \cite{feng_review_2022}, used otherwise in the following results.

\texttt{FLARE} was used in this work to create a magnetic mesh for the equilibrium magnetic field and for field line diffusion approximations of heat loads on the divertor. Once the divertor plate files have been created by the divertor-making algorithm, they are passed to \texttt{FLARE}, which converts the plate files from Kisslinger format to a computational mesh. \texttt{FLARE} evaluates heat flux on the divertor by computing the intersection of field line trajectories on the magnetic mesh with the computational mesh of the divertor.

The simulation parameters, unless otherwise specified, are shown in Table~\ref{tab:sim_params}.
\begin{table}
    \centering
    \begin{tabular}{ll}
        \hline
        \textbf{Parameter} & \textbf{Value} \\
        \hline
        $n_\text{particles}$ & $10^6$ \\
        $n_0$ & $10^{19}~\text{m}^{-3}$ \\
        $T_i$ & $50~\text{eV}$ \\
        $\chi_\perp$ & $0.1~\text{m}^2/\text{s}$ \\
        $P_\text{SOL}$ & $8~\text{MW}$ \\
        \hline
    \end{tabular}
    \caption{Parameters for \texttt{FLARE} simulations.}
    \label{tab:sim_params}
\end{table}
The cross-field diffusion coefficient $\chi_\perp=0.1~\text{m}^2/\text{s}$ is justified by assuming that the perpendicular ion transport will be gyro-Bohm, and that $\chi\sim1/B$ \cite{manfredi_gyrobohm_1997}, assuming a fixed edge temperature for simplicity. We also assume $\chi_e/\chi_i\approx1$ \cite{wappl_experimental_2025}. W7-X discharges have $B=2.5~\text{T}$ and the equilibrium used in this work has $B=9~\text{T}$, so we expect $\chi$ to be reduced by a factor of $3.6$. Using W7-X measurements that report an edge $\chi_e\approx0.5~\text{m}^2/\text{s}$ \cite{weir_heat_2021}, in this equilibrium we assume $\chi_i\approx0.5/3.6\approx0.1~\text{m}^2/\text{s}$. It should be noted that this value is still a fairly conservative estimate, as larger values would increase the heat flux spread on the divertor plate and reduce $q_{\text{peak}}$, as shown in Sec.~\ref{subsec:divertor_robustness}. \texttt{FLARE} returns $\text{\# particles hit}/\text{m}^2$ across the computational domain. This is converted to a heat flux by multiplying by the assumed power entering the scrape-off-layer, $P_\text{SOL}$, per half field period \cite{bader_power_2025}.

\subsection{Divertor heat flux performance metric}\label{subsec:divertor_performance}

\begin{figure}
    \centering
    \includegraphics[width=0.75\linewidth]{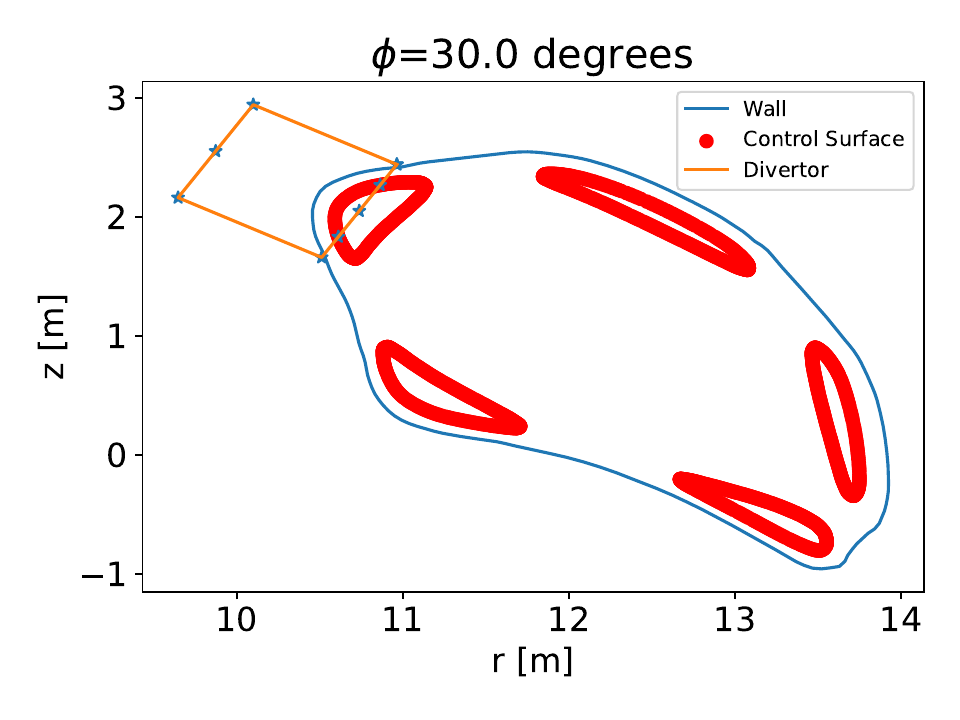}
    \caption{The cross section of the computational boundary  of the divertor end-plate with respect to the island geometry and vessel wall. The vessel wall and end-plates are included in the \texttt{FLARE} simulation to compute the power fraction coming from the core that misses the plasma-facing divertor plates. The blue stars indicate the coordinates of the end-plate used in the divertor end-plate file.}
    \label{fig:divertor_cross_section}
\end{figure}

A metric is needed to compare the performance of different divertors produced by the divertor-making algorithm. We use a cost function for this purpose, as it will be used to guide optimization of divertors later on in Sec.~\ref{subsec:optimization}. We use a relatively simple cost function that only targets the divertor performance with respect to heat flux, given by
\begin{equation}\label{eqn:cost_func}
    J(\theta_L, \theta_R)=W_1~q_\text{peak}+W_2~(100-P_\text{captured}),
\end{equation}
where $\theta_L$ and $\theta_R$ are initial conditions for the divertor-making algorithm, $W_1$ and $W_2$ are user determined weights, $q_{\text{peak}}=\max{(q)}$ is the maximum heat flux, and $P_\text{captured}$ is the non-radiated power fraction that hits the divertor plates, in percent. This is computed as the fraction of fieldlines that hit the divertor plates, as some fieldlines may miss the divertor plates by intercepting the vessel wall or the sides of the divertor plate that are not meant to interact with plasma. Until recently, \texttt{FLARE} did not discriminate between field line hits on the plasma-facing side or the vessel facing side of the divertor plate. As a result, a computational domain was constructed to make a rectangular prism shape that models the sides of the divertor plates. This is implemented by ensuring that the divertor points saved in Kisslinger format create a closed structure, and by using additional divertor plate files that extend a fraction of a degree past $\phi_{\text{start}}$ and $\phi_\text{end}$ and create an end-plate extending radially away from the island towards the wall, as shown in Fig.~\ref{fig:divertor_cross_section}. These extra computational boundaries are only used to help find good divertor solutions by making it possible to calculate $P_\text{captured}$, and are not a part of the final divertor design. 

The cost function combines the objectives of minimizing the peak heat flux while ensuring that the divertor captures the majority of the non-radiated power from the core plasma.

\section{Results}\label{sec:results}

\begin{figure}
    \centering
    \includegraphics[width=0.75\linewidth]{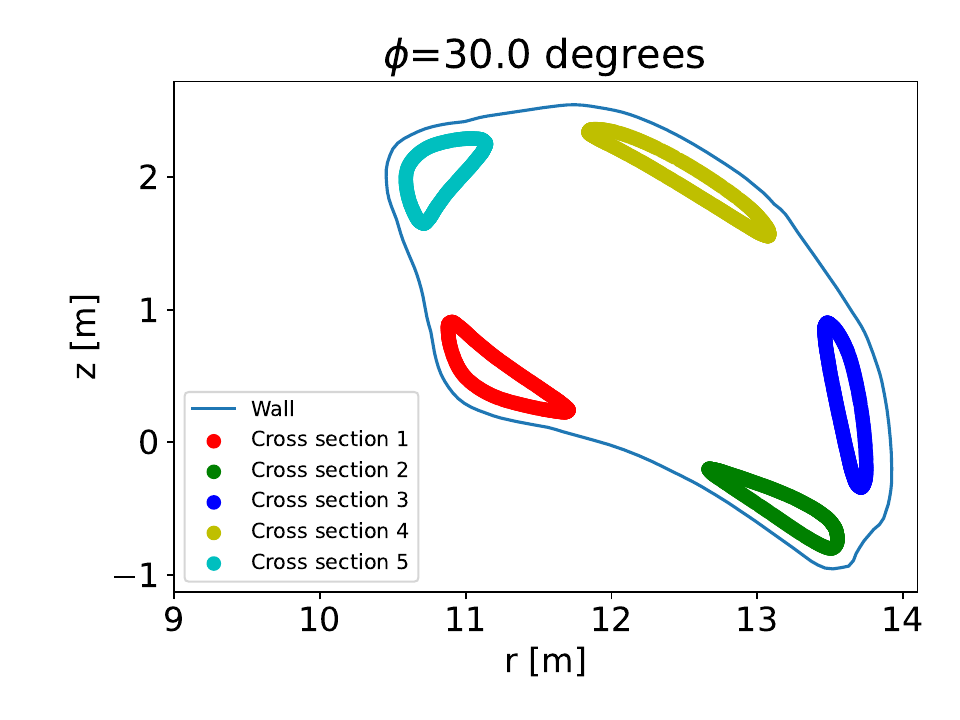}
    \caption{A poloidal cross section of the island chain at $\phi=30^\circ$. The results section presents a detailed analysis of creating divertors for cross section 5, before comparing the different cross sections with each other with respect to divertor compatibility.}
    \label{fig:island_cross_section}
\end{figure}

This section is organized first into the validation of the Bayesian optimization against a parameter scan by looking at divertors on a single island of the island chain. The divertor algorithm parameters are fixed to be $\phi_\text{start}=15^\circ,\phi_\text{init}=30^\circ, \phi_\text{end}=45^\circ$, and $\Delta\phi=0.5^\circ$. The Bayesian optimization finds a near-identical divertor to the grid scan, and validates the use of Bayesian optimization to find divertors. Then we discuss the application of Bayesian optimization to explore the larger equilibrium parameter space, which may be used as an assessment of an equilibrium's compatibility with island divertors. This is done by performing optimizations over all the islands shown in Fig.~\ref{fig:island_cross_section}.

\subsection{Divertor grid scan over starting angles}\label{subsec:grid_scan}
\begin{figure}
    \centering
    \includegraphics[width=\linewidth]{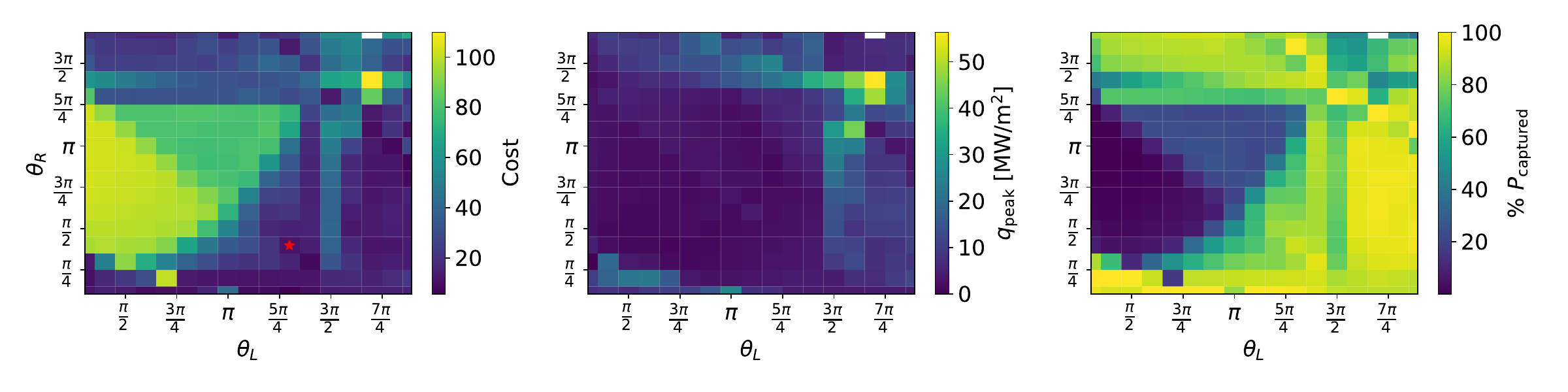}
    \caption{Cost function landscape for 441 divertors chosen from starting angles between $0$ and $2\pi$, as well as the individual terms $q_\text{peak}$ and $P_\text{captured}$. The axes are scaled to capture the region of parameter space where the divertor-making algorithm completed successfully. The white area represents $(\theta_L, \theta_R)$ initial conditions that did not produce a divertor because the divertor-making algorithm crashed. Regions of very high cost that seem discontinuous, such as in the upper right or lower left corners, are a result of leading edges on the end-plates, as shown in spikes of $q_\text{peak}$. The best performing divertor at $(\theta_L, \theta_R)=(4.08, 1.25)$ is denoted by the red star.}
    \label{fig:cost_function_landscape}
\end{figure}

Since we have simplified the divertor-making algorithm to two parameters, we can visualize the entire parameter space with a contour plot. We employ a simple two-dimensional scan across the divertor parameter space. We create 441 pairs of starting angles by discretizing the control surface of island 5 over the range $[0, 2\pi)$ with 21 points and choosing all combinations.
Each set of angles $(\theta_L, \theta_R)$ on the $\phi_\text{init}$ plane is used as initial conditions to create divertors. In the case where $\theta_L=\theta_R$, the plates have zero poloidal thickness for half of a degree before $\theta_L$ and $\theta_R$ are modified by the algorithm to no longer be equal as the divertor algorithm progresses, and this does not affect the divertors' ability to capture heat. This effect was also observed when $\theta_L$ and $\theta_R$ cross over each other as the divertor algorithm progresses, creating bowtie-shaped divertors, but once again this does not affect the divertors' ability to capture heat. This came about because we avoided restricting the parameter space of $(\theta_L, \theta_R)$ initial conditions, and could be remedied with further postprocessing to fill in the missing gap. Heat loads are evaluated on every divertor using the \texttt{heat\_load\_proxy} task in \texttt{FLARE}, which performs a field line diffusion simulation. The results of the grid scan are shown in Fig.~\ref{fig:cost_function_landscape}. The white area represents $(\theta_L, \theta_R)$ initial conditions that did not produce a divertor because the divertor-making algorithm crashed. The reason for these crashes was that a numerical root solver used to identify the point on the island spline for the correct strike angle could not converge. Outside of the axes limits in Fig.~\ref{fig:cost_function_landscape}, the divertor algorithm was not able to produce a viable divertor.  Despite this, large regions of viable parameter space still exist. Out of the 441 divertor initial conditions chosen, 288 completed successfully, leading to a $65\%$ success rate. However, the cost function landscape does have areas of non-differentiability. In addition, it is non-convex. If more parameters were added to the divertor-making algorithm, the cost function landscape would increase in dimensionality and complexity. The computational cost of such a scan is $\sim5$ to $10$ minutes per divertor using $128$ cores on one Perlmutter AMD EPYC 7763 CPU node.

The lowest cost divertor from the grid scan has initial conditions $(\theta_L, \theta_R)=(4.08, 1.25)$, $q_{\text{peak}}=2.915~\text{MW}/\text{m}^2$, and a $P_\text{captured}$ of 91.94\%, and is denoted by the red star in Fig.~\ref{fig:cost_function_landscape}.


\subsection{Bayesian Optimization}\label{subsec:optimization}

Although the computational efficiency of \texttt{FLARE} is sufficient to run large grid scans and quickly evaluate a given divertor configuration with a reduced physics model, the computational expense is non-negligible, even with only two parameters to vary. Adding more degrees of freedom to the divertor-making algorithm would exacerbate this problem. We seek a computationally efficient way to search the algorithm parameter space and find a satisfactory divertor configuration. Ideally, we find the divertor that minimizes Eqn.~\ref{eqn:cost_func}. Bayesian optimization \cite{frazier_2018} is a popular optimization method for computationally expensive problems where gradients are not available. In our case, there are no gradients available using auto-differentiation given the statistical nature of \texttt{FLARE} simulations. Computing gradients using finite differences is computationally prohibitive, and gradient-based methods could struggle with the non-convex cost function landscape in Fig.~\ref{fig:cost_function_landscape}, especially if more optimization variables were added. Bayesian optimization, on the other hand, is a global optimization method that may work well on these problems. 

Bayesian optimization relies on statistical methods by assuming an underlying Gaussian process that models the cost function. This Gaussian process is assigned a covariance function, or kernel, which determines how observations in parameter space are correlated with each other. After making initial, random observations in parameter space, Bayesian optimization methods make their next observations based on an acquisition function. The acquisition function takes into account the mean of the Gaussian process and its standard deviation to balance exploration of parameter space and finding the optimal solution. On every step of the optimization, the Gaussian process is updated, providing more information to make the next guess. After a few dozen iterations, the resulting Gaussian process can accurately model regions of parameter space and find the global optimum \cite{jones_efficient_1998}.

The open-source \texttt{BayesianOptimization} Python package \cite{nogueira_bayesian_2020} is applied to the divertor shape optimization problem. The Gaussian process was set up to use a Mat$\acute{\text{e}}$rn kernel \cite{frazier_2018}, given by 
$$
    k(x_i, x_j)=\frac{1}{\Gamma(\nu)2^{\nu-1}}\left(\frac{\sqrt{2\nu}}{l}d(x_i, x_j)\right)^\nu K_\nu\left(\frac{\sqrt{2\nu}}{l}d(x_i, x_j)\right),
$$
where $d(\cdot,\cdot)$ is the Euclidean distance, $K_\nu(\cdot)$ is a modified Bessel function of order $\nu$ and $\Gamma(\cdot)$ is the gamma function. The parameter $\nu$ controls the smoothness of the kernel \cite{rasmussen_2010}, and was set to $2.5$, a standard choice in Bayesian optimization \cite{shahriari_bo_review_2016}. A soft barrier penalty of $200$ was found to work well in steering the model away from pairs of $(\theta_L, \theta_R)$ that fail to produce a viable divertor. Because of this, the length scale of the Gaussian process is bounded between $0.1\times$ and $1.5\times$ the range of the domain in $\theta_L$ and $\theta_R$. This ensures that the length scale does not collapse to near zero when imposing the soft penalty, which was noticed to be an issue in optimization. The optimizer uses an Expected Improvement \cite{zhan_expected_2020} acquisition function equipped with a hyperparameter $\xi$ that determines the relative weighting between exploitation and exploration of the cost function. This is in principle another hyperparameter that needs to be tuned. We tested $\xi$ values of 0.005, 0.01, and 0.05. The best results were achieved using $\xi=0.01$, and are presented below.

\begin{figure}
    \centering
        \includegraphics[width=0.7\linewidth]{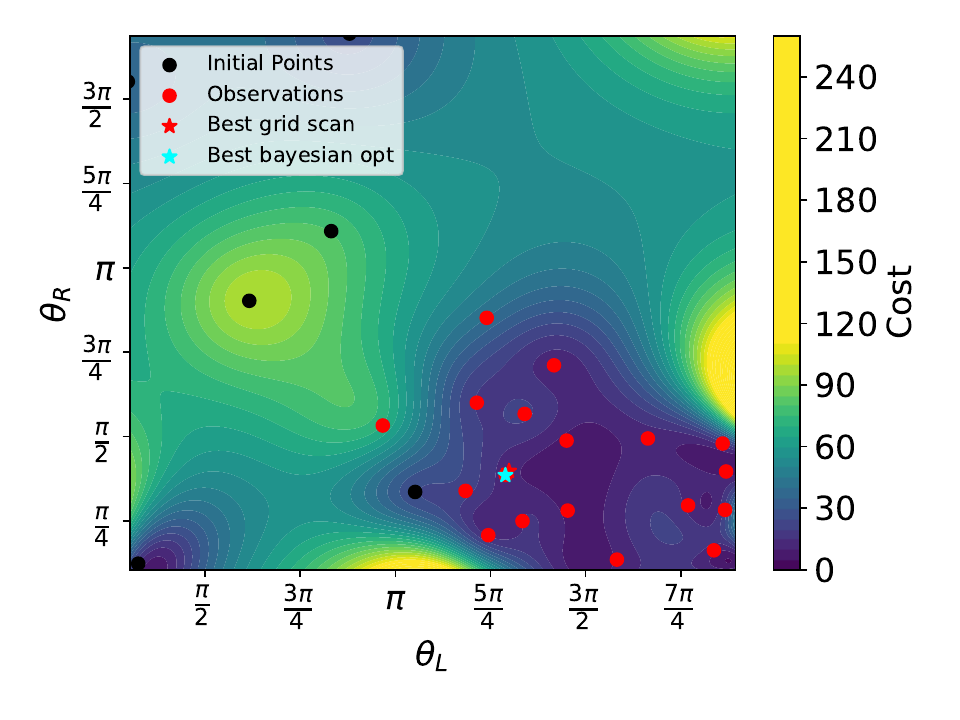}
    \caption{Cost function landscape as learned by a Bayesian optimization run. In black are the initial points used in the optimization, and in red the subsequent observations taken by the optimizer. The optimizer finds the large valley around $(\theta_L, \theta_R)=(4.5, 1.5)$. The optimal divertor from the Bayesian optimization is denoted in cyan, and is very close to the optimal divertor found by the grid scan.}
    \label{fig:bayesopt}
\end{figure}

The optimization was initialized with 10 randomized initial steps and with 100 exploratory steps guided by the acquisition function. Fig.~\ref{fig:bayesopt} shows the cost function landscape as learned by Bayesian optimization after only 25 exploratory steps. The majority of points are concentrated in the lower right corner, a valley in parameter space that is present in Fig.~\ref{fig:cost_function_landscape} as well. At this point in the optimization, we find a divertor that is near-identical to the best one from the grid scan. The optimized divertor was found at $(\theta_L, \theta_R)=(4.05, 1.21)$, with $q_{\text{peak}}=3~ \text{MW}/\text{m}^2$ and a $P_\text{captured}$ of $92.49\%$, and is denoted by the cyan star in Fig.~\ref{fig:bayesopt}. However, in general more exploratory steps are required to find the local minimum, and this specific case is relatively fortunate

The Bayesian optimization achieves practically the same result as the grid scan but with a $75\%$ reduction in the number of divertors simulated. This result highlights the power of Bayesian approaches to expensive, black-box, and non-convex optimization problems. We expect that the savings would increase even more if the divertor-making algorithm parameter space were larger. 

The optimizer does not exactly recreate the cost function landscape in Fig.~\ref{fig:cost_function_landscape}, but there is no need to.  The power in the optimization is to rapidly identify good regions of parameter space that should be further explored, either by further optimizations with a smaller domain, or more focused parameter grid scans. 

\begin{figure}
    \centering
    \includegraphics[width=0.7\linewidth]{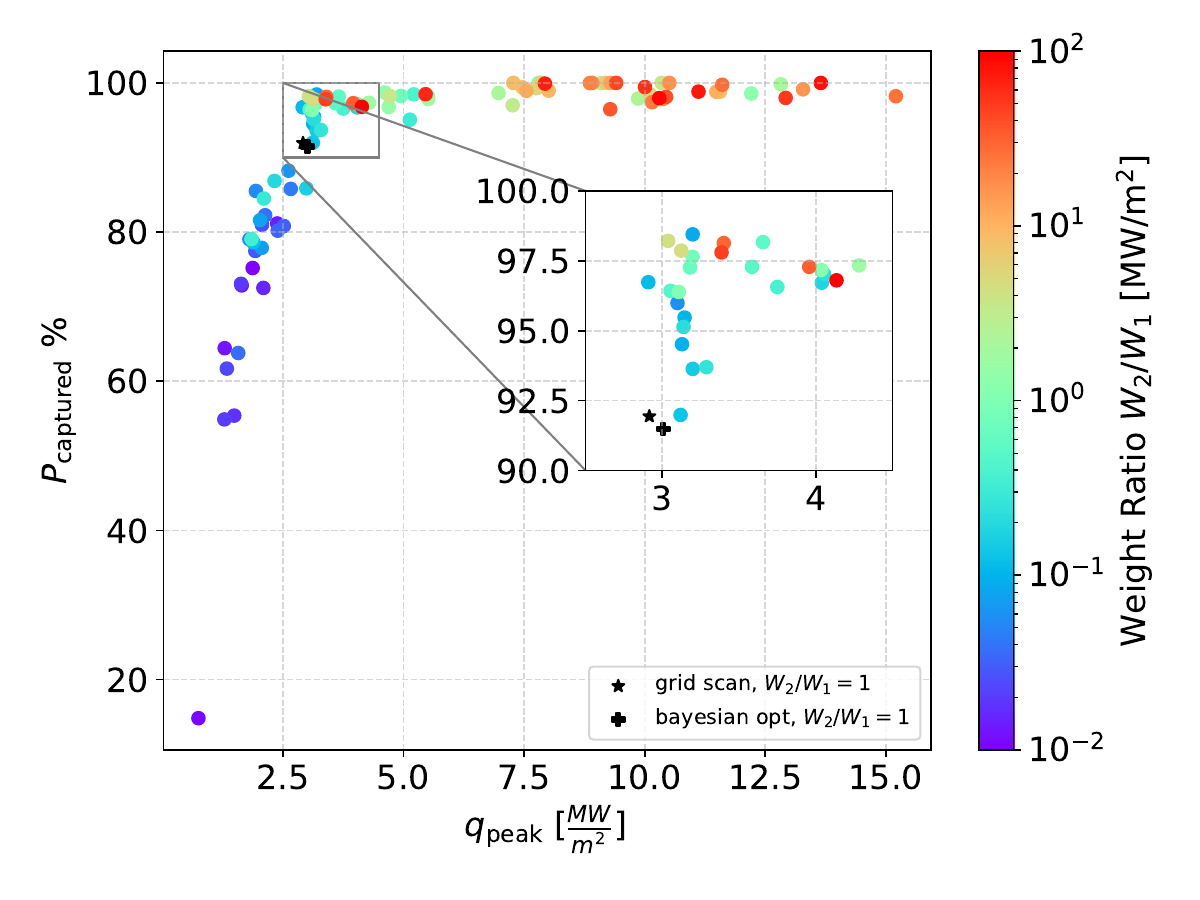}
    \caption{The Pareto frontier represents the best possible divertors when considering the trade-off between the two terms in the cost function, $q_\text{peak}$ and $P_\text{captured}$. The results from the best divertors from the grid scan and the Bayesian optimization are also shown, and lie on the Pareto frontier.}
    \label{fig:pareto_frontier}
\end{figure}

A concept useful in optimization is the Pareto frontier. The Pareto frontier represents the trade-off between objectives in an optimization problem. In our case, the trade-off is the minimization of peak heat flux and ensuring that $P_\text{captured}$ remains high. We vary the weight ratio of the two terms in the cost function, $W_2/W_1$, and run an optimization with the cost function given by Eqn.~\ref{eqn:cost_func} with the new weight ratio. We ran 100 optimizations with logarithmically spaced weight ratios from $10^{-2}$ to $10^2$, with 10 initial sampling steps and 20 exploratory steps. The resulting Pareto frontier formed by the lowest cost divertors in each case is shown in Fig.~\ref{fig:pareto_frontier}. Solutions along this Pareto frontier represent the best possible trade-off between $q_{\text{peak}}$ and $P_\text{captured}$.  Fig.~\ref{fig:pareto_frontier} shows that to get to a $P_\text{captured}$ greater than 90\%, $q_{\text{peak}}\gtrsim3~\text{MW}/\text{m}^2$. There is some scatter in points with similar weight ratios, where some may have a very low $q_{\text{peak}}$ while others have a $q_\text{peak}>10~\text{MW}/\text{m}^2$, and occurs mainly when $W_2/W_1>1$. This is attributed to the fact that not enough exploratory steps were taken in these optimizations to consistently locate the optimal divertor. When $W_2/W_1>1$, the cost function landscape structure is dominated by the power fraction captured, so that the valleys have large variation in $q_\text{peak}$. The majority of optimizations in the Pareto scan do find these valleys in their respective cost function landscapes, however, as only 20 exploratory steps are used, as opposed to 100 in the earlier optimization highlighted, the optimum of the valley is not always found. Thus, the solutions in the valleys exhibit variation in $q_\text{peak}$. Increasing the number of steps or further hyperparameter tuning could improve the optimization behavior. Additionally, in Fig.~\ref{fig:cost_function_landscape} and Fig.~\ref{fig:bayesopt}, the cost function landscape is seen to be non-convex. Since the optimizations have randomized initial conditions, there is no guarantee that two optimizations with similar weight ratios will converge to the same optimum, which could contribute to the spread. 

We also analyze the Pareto frontier optimizations in terms of how many found a divertor solution that meets the physics criteria of $q_\text{peak}< 5~\text{MW}/\text{m}^2$ and $P_\text{captured}>90\%$. Of the 100 optimizations, 86 of them found a divertor meeting these requirements in under 30 optimization steps. The solutions that meet these physics requirements are not necessarily the points shown in Fig.~\ref{fig:pareto_frontier}.

\subsection{Heat Flux Profiles}

\begin{figure}
    \centering
    \includegraphics[width=0.7\linewidth]{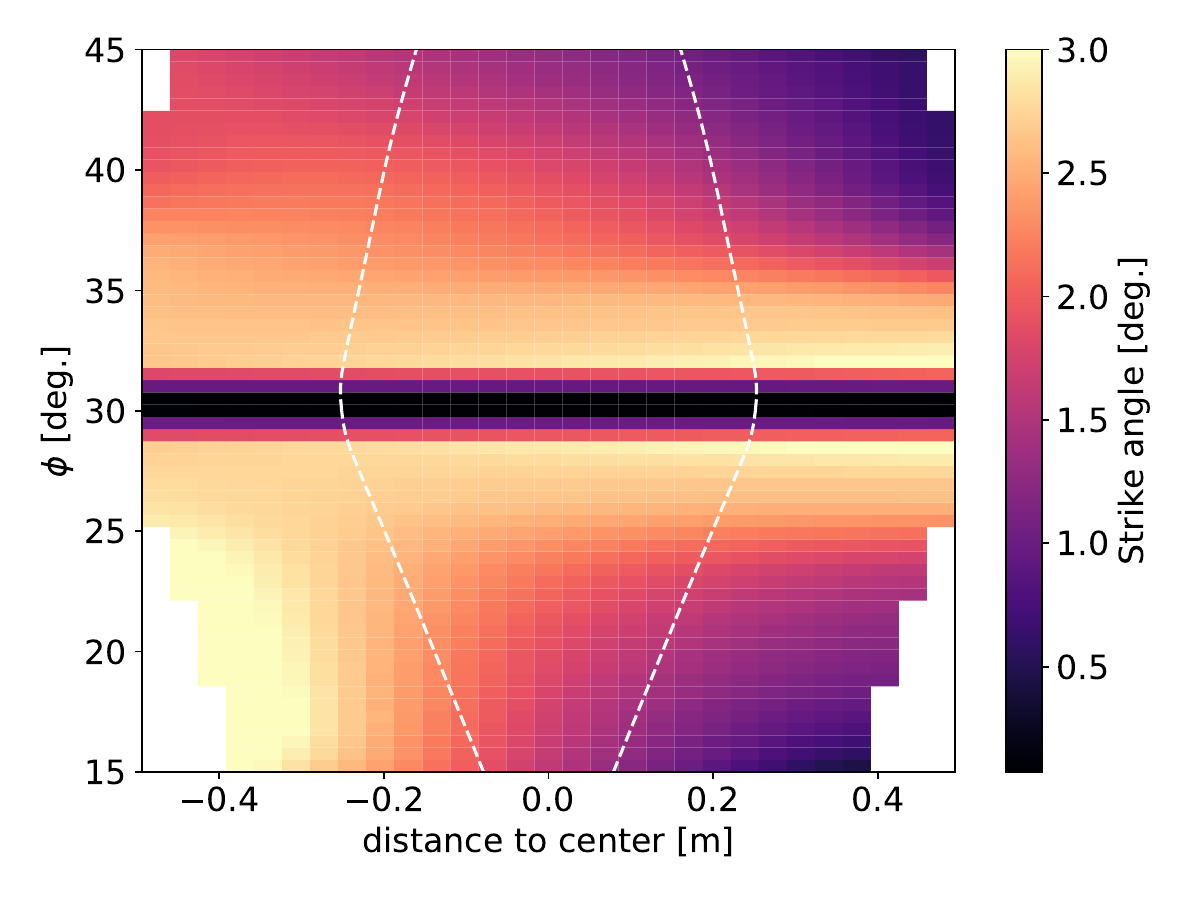}
    \caption{Strike angles on the divertor plate for a divertor created with $(\theta_L, \theta_R)=(4.05, 1.21)$. The strike angle varies primarily between $1^\circ$ and $3^\circ$. The ``round-off" where the strike angle decreases linearly from $3^\circ$ to $0.1^\circ$ can be seen near $\phi=30^\circ$.  The white dotted lines are where the control surface intersects the divertor plate.}
    \label{fig:strike_angle}
\end{figure}

\begin{figure}
    \centering
    \includegraphics[width=0.7\linewidth]{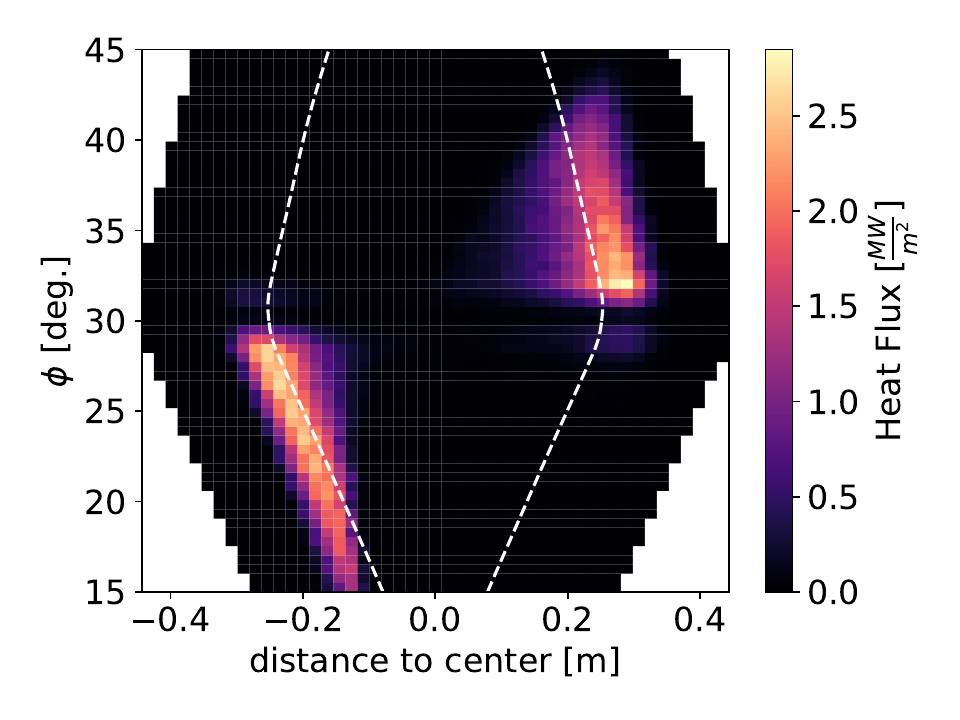}
    \caption{Heat flux $q$ pattern on the divertor plate for a divertor created with $(\theta_L, \theta_R)=(4.05, 1.21)$. The black area represents the divertor computational domain. The peak heat flux is far below the $10~\text{MW}/\text{m}^2$ limit. The white dotted lines are where the control surface intersects the divertor plate. The misalignment is due to the control surface imperfectly representing the separatrix.}
    \label{fig:heat_flux}
\end{figure}

\begin{figure}
    \centering
    \includegraphics[width=\linewidth]{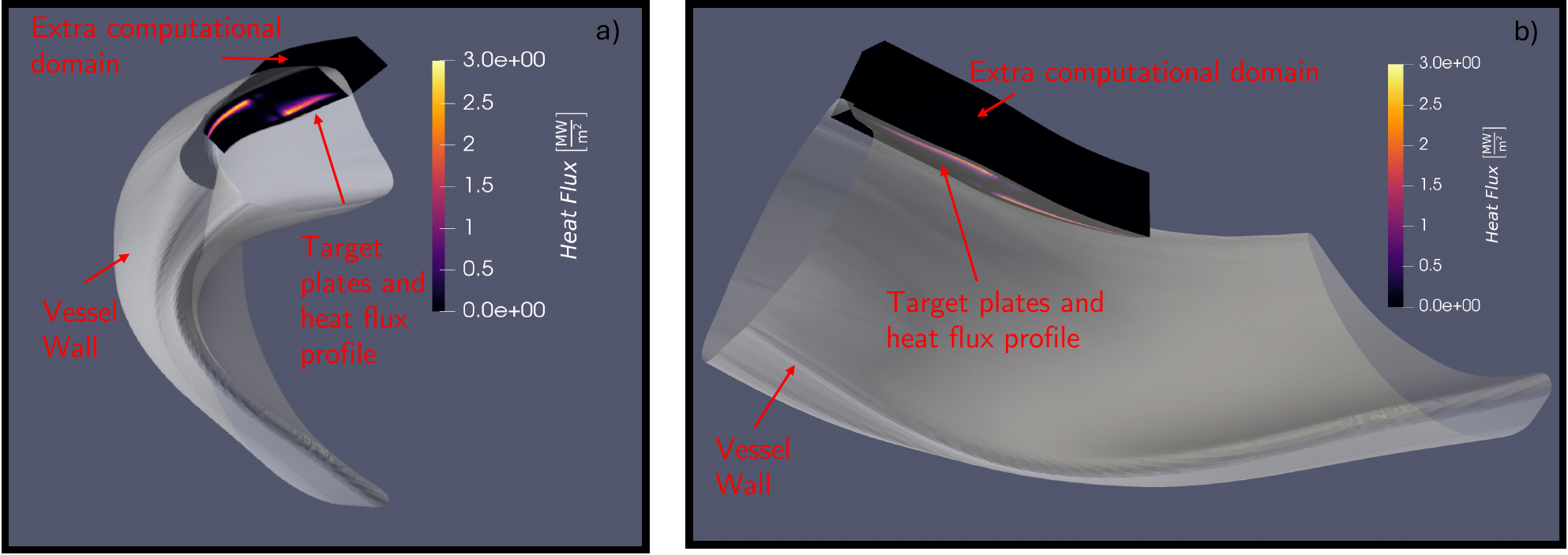}
    \caption{Heat flux $q$ pattern on the divertor plate visualized in 3D. The vessel wall is shown in gray. The computational domain for the divertor is shown in black. The closing of the divertor structure past the domain of the vessel wall to avoid erroneous field line hits on the backside of the divertor can also be seen. Subplots a) and b) show the same simulation, but from different angles to showcase that the divertor is field aligned.}
    \label{fig:3d_heat_flux}
\end{figure}

In this section, we present the performance of the best optimized divertor in terms of the heat flux profiles on the divertor plate. The enforcement of a maximum strike angle is shown in Fig.~\ref{fig:strike_angle}. The heat flux profile for the divertor described in Sec.~\ref{subsec:optimization} is shown in Fig.~\ref{fig:heat_flux}. A 3D visualization is shown in Fig.~\ref{fig:3d_heat_flux}. In Fig.~\ref{fig:strike_angle} and Fig.~\ref{fig:heat_flux}, the white area represents regions with no divertor. The x-axis is parametrized in terms of distance from the plate center, as the divertor-making algorithm produces divertors that change width as the divertor plates move away from the core plasma further into the island.

\subsection{Heat flux width scaling and divertor robustness}\label{subsec:divertor_robustness}

The optimization was performed with a single set of plasma parameters. However, a good divertor should ideally work across a range of plasma parameters, which may be seen, for example, in the transition from attachment to detachment \cite{peterson_investigation_2025}, or during plasma start-up \cite{zhou_equilibrium_2022}. To test the robustness of the divertor to different plasma conditions, the artificial cross-field diffusion coefficient $D$ is varied for the optimized divertor shown in Fig.~\ref{fig:3d_heat_flux}. A more detailed explanation of the artificial cross-field diffusion coefficient can be found in \citet{frerichs_magnetic_2025}.

\begin{figure}
    \centering
    \includegraphics[width=0.7\linewidth]{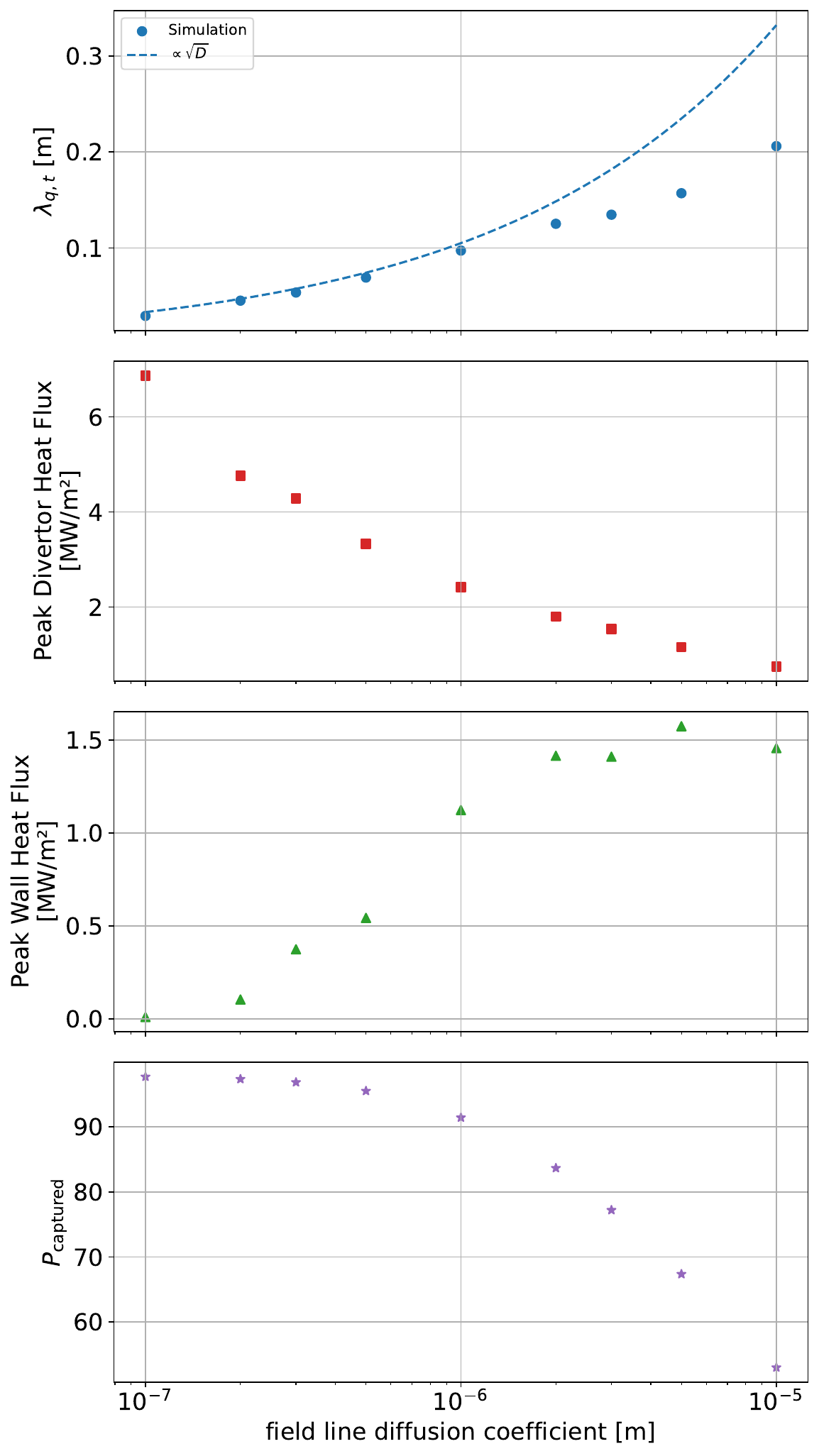}
    \caption{Scaling of $\lambda_{\text{q,t}}$ with respect to the cross-field diffusion coefficient. $q_\text{peak}$ decreases as the cross-field diffusion coefficient increases, as the incident power is spread out over a larger area. The scaling from \citet{feng_review_2022} does not hold for high diffusivities as fieldlines hit the vessel wall rather than the divertor plate, demonstrated by the increasing wall heat flux and the decreasing $P_\text{captured}$.}
    \label{fig:lambda_q_scaling}
\end{figure}

The measure of robustness used is the expected spreading of the heat flux over the target plate. The heat flux width on the target divertor plate is denoted as $\lambda_{\text{q,t}}$. Assuming a scrape-off layer with a constant $L_c$, the heat flux will spread out over a length scale of \cite{feng_review_2022}
\begin{equation}
    \lambda_{\text{q,t}}\sim\sqrt{\frac{DL_c}{C_s}}.
    \label{eqn:lambda_q_scaling}
\end{equation}
By keeping $C_s$ fixed in the scan of diffusion coefficients, we expect $\lambda_\text{q,t}\sim\sqrt{D}$. We measure $\lambda_{\text{q,t}}$ from simulation data following the procedure in \citet{bader_power_2025}. For every toroidal angle of the divertor plate, we extract the 1D heat flux profile as a function of the distance to the center of the plate. To this distribution we fit a Gaussian distribution using \texttt{scipy.optimize.curve\_fit}. A good proxy for $\lambda_{\text{q,t}}$ is an average over the toroidal angles of the full widths at half maximum (FWHM) of the heat flux profiles. The formula is given by 
\begin{equation}
    \lambda_{\text{q,t}}=\frac{1}{N}\sum_{i=1}^{N}(\text{FWHM})_i\frac{\max(q)_i}{\max(q)},
\end{equation}
where the index $i$ represents the profiles at toroidal angle $\phi=\phi_{\text{init}} +i\times\Delta{\phi}$, and $N=|(\phi_{\text{init}}-\phi_\text{stop})|/\Delta\phi$.
The scaling with respect to $D$ is shown in Fig.~\ref{fig:lambda_q_scaling}. Good agreement is seen for low diffusivities, but the scaling no longer agrees well for diffusivities above $3
\times10^{-6}~\text{m}$. The disagreement of the $\lambda_{\text{q,t}}$ scaling can be explained by looking at the percentage of particles that hit the vessel wall rather than the divertor plate, which increases with increasing diffusion coefficient. Having a significant fraction of particles striking the wall means that the divertor is not capturing all the heat flux, and the scaling in Eqn.~\ref{eqn:lambda_q_scaling} no longer holds. In the set of simulations run in this work, the results deviate from the scaling in Eqn.~\ref{eqn:lambda_q_scaling} once the wall hit percentage is above 5\%.

\subsection{Evaluation of equilibrium compatibility with island divertors}
The divertor algorithm described above is useful in that it is a general method that can attempt to build divertors for any island geometry with minimal user input or initial guesses, which is valuable for a validation of an equilibrium's compatibility for island divertors as a heat flux solution. We have demonstrated that Bayesian optimization can successfully find divertor geometries that meet heat flux requirements, and now we apply the framework to evaluate the larger equilibrium parameter space for island divertor viability. A natural question when designing an island divertor is: at what toroidal and poloidal locations should the islands be intersected? We try to answer this question by assessing the viability of all five island cross sections of the equilibrium, shown in Fig.~\ref{fig:island_cross_section}. In this study, the toroidal extent of the divertor is fixed at $30^\circ$. Since the equilibrium has four field periods and is stellarator symmetric, $\phi_\text{init}$ can be at minimum at $\phi=15^\circ$ and at maximum $\phi=30^\circ$. Optimizations are performed at four different toroidal positions for $\phi_\text{init}$ for all five cross sections. We avoid adding $\phi_\text{init}$ as an optimization variable as it is a discrete variable in our representation of the control surface. Future work could use an interpolant to make $\phi_\text{init}$ continuous, allowing it to be an optimization parameter.

For every single cross section-angle pair, 20 different Bayesian optimizations are performed with weight ratios log-uniformly distributed between $10^{-1}$ and $10^{1}$. This range of weights was picked based off the Pareto frontier results shown in Fig.~\ref{fig:pareto_frontier}, where it can be seen that weights between $10^{-1}$ and $10^1$ capture the range of weights where better divertor solutions exist. Every optimization runs for 50 iterations, which is about twice as long as the optimization highlighted in the previous section.

If any divertor from the 20 optimizations meets the criteria of $q_\text{peak} < 5 ~\text{MW}/\text{m}^2$ and $P_\text{captured}>90\%$, then the optimization is considered successful. A limit of $q_\text{peak} < 5 ~\text{MW}/\text{m}^2$ is conservative with respect to the $10~\text{MW}/\text{m}^2$ material limit, and is meant to provide a factor of safety of two, given the assumed scrape-off-layer power and the use of a simpler heat flux model.

The results of the study are shown in Table~\ref{tab:equilibrium_scan}. It is apparent that for this equilibrium, cross section 5 is the best candidate for an island divertor as the optimization is successful for every toroidal location of the divertor. 

\begin{figure}
    \centering
    \includegraphics[width=0.75\linewidth]{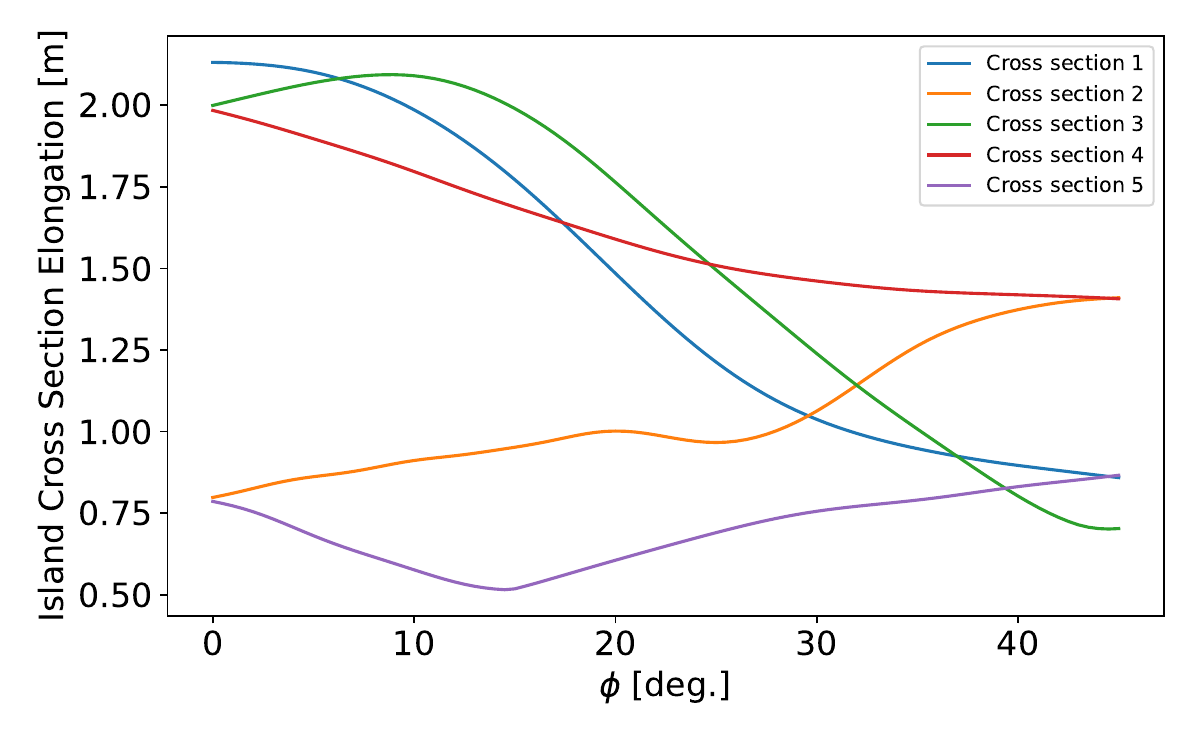}
    \caption{The elongation of every island cross section as a function of toroidal angle. The elongation is measured as the maximum between any two points on the cross section of the control surface. Cross section 5 has the lowest elongation, making it more suitable for finding an island divertor that is compatible with the heat flux and power fraction captured requirements. The optimization struggles with elongated cross sections whose shape changes significantly as a function of toroidal angle.}
    \label{fig:island_elongation}
\end{figure}

The algorithm struggles to find divertors for other cross sections, which are too elongated. Elongated cross sections have a magnetic structure that changes rapidly with respect to the toroidal angle $\phi$. We confirm this by parameterizing the change in the island cross section structure by computing its elongation. Fig.~\ref{fig:island_elongation} shows that cross section 5 has both the smallest and most constant elongation, which is more suitable for our divertor generation algorithm. This line of reasoning also supports the success at cross section 2, $\phi_\text{init}=15^\circ$, which also has a lower and more constant elongation than for the majority of the other cases. The large variation in elongation leads to numerical issues in the algorithm, related to the root solver used being sensitive as the island geometry changes rapidly, as it does in elongated cross sections. This leads to large inaccessible regions of parameter space, and difficulty for the Bayesian optimization to find a divertor that meets the heat flux and power fraction captured requirements. Optimization struggles because algorithm failures result in cost barriers, making the optimization landscape less smooth. Elongated cross sections may benefit from a more complex divertor geometry generation algorithm, but it should also be noted that engineering and manufacturing challenges could arise for a divertor on an elongated island cross section which changes orientation.

This type of equilibrium study highlights the power of coupling a divertor creation algorithm with Bayesian optimization to efficiently characterize an equilibrium's compatibility with an island divertor. If a scan over the initial $(\theta_L, \theta_R)$ angles were performed for each 
divertor position instead of applying the Bayesian optimization, the computational cost would increase from dozens of CPU hours to thousands. Quickly knowing where an island divertor can be placed is valuable information for the overall reactor design process. Further optimizations to find the best divertor position and geometry can then be performed.

\begin{table}[htbp]
    \centering
    \caption{Results of Bayesian optimizations searching for compatible divertors in the reference equilibrium. Checkmarks indicate that a divertor with $q_\text{peak}<5~\text{MW/m}^2$ and $P_\text{captured}>90\%$ was found, while crosses indicate that a suitable divertor was not found.}
    \label{tab:equilibrium_scan}
    \begin{tabular}{cccccc}
    & \multicolumn{5}{c}{\textbf{Cross section}} \\
    \cmidrule{2-6}
    $\mathbf{\phi}_\text{init}$ & \textbf{1} & \textbf{2} & \textbf{3} & \textbf{4} & \textbf{5} \\
    \midrule
    15 & $\failure$ & $\success$ & $\failure$ & $\failure$ & $\success$ \\
    20 & $\failure$ & $\failure$ & $\failure$ & $\failure$ & $\success$ \\
    25 & $\failure$ & $\failure$ & $\failure$ & $\failure$ & $\success$ \\
    30 & $\failure$ & $\failure$ & $\failure$ & $\failure$ & $\success$ \\
    \bottomrule
    \end{tabular}
\end{table}

\section{Discussion and conclusions}\label{sec:conclusions}

This work presents an automated design and optimization of stellarator island divertors in fixed magnetic equilibria to help inform heat flux solutions in stellarator FPPs. The optimization minimizes the peak heat flux incident on the divertor plate by changing the shape of the divertor with respect to the shape of the magnetic island boundary. The optimization finds divertors that stay below the material limit of $10~\text{MW}/\text{m}^2$ even under a conservative radiative fraction estimate of 60\%, which is about 30 percentage points less than other radiative fraction estimates for stellarator FPPs \cite{hegna_2025, swanson_overview_2026}.

The optimization here is a proof-of-principle application to the boundary of FPPs, and there are many additions that can be made to support island divertor design for the stellarator community. First, this work only attempts to reduce the peak heat loads incident on the divertor plate. This work does not consider engineering constraints such as space for cooling and support structures, and manufacturing and alignment tolerances \cite{lore_design_2014}. Future work should include terms in the cost function to better address these constraints. 

This applies as well to critical physics objectives that must be considered for any island divertor. Examples of these are particle exhaust, particle retention, and maintaining a high neutral pressure. The lack of a neutral physics model limits the conclusions one can draw about the resulting divertors' performance in a reactor-like scenario for particle exhaust. Open island divertors have poor neutral retention and therefore low pumping efficiencies \cite{sunn_pedersen_first_2019}. Neutral particle emissions from the divertor plate follow the cosine distribution, and the neutral flux would be directed normal to the divertor into the plasma, resulting in low neutral retention. Since the divertors shown in this paper are all open divertors, their neutral performance would most likely follow the results of W7-X. If pumping gaps were placed in similar orientations to W7-X, it is expected that wide strikeline profiles and a heat flux of about 5 $\text{MW}/\text{m}^2$ would provide similar pumping efficiencies \cite{bader_power_2025}. Open stellarator divertors could benefit from further studies of toroidally oriented pumping gaps in private flux regions \cite{Moscheni2026} to help mitigate this problem. 

The most natural first step would be to add neutral and re-ionization models into the optimization scheme, in a way similar to \cite{frerichs_2026}. A similar consideration should be made for impurity transport. Additionally, since the computational model only resolves one half field period, there could be shadowing between the real divertor plates, which would be repeated 8 times. Furthermore, the existence of private flux regions and target shadowed regions, and the differences in heat flux transport in them \cite{Kharwandikar2026}, are also not included in this analysis. These could be studied with codes that resolve particle transport with higher fidelity.

There is also room to add complexity to the divertor-making algorithm. Some free parameters that were held fixed in this work but could be optimization parameters include the maximum strike angle $\alpha$ and the extent of the divertor with respect to the toroidal angle.

Eventually, divertor optimization tools should be integrated with the rest of the stellarator optimization suite. One could either evaluate the performance of equilibria with respect to a divertor using field line diffusion codes like \texttt{FLARE} within the optimization loop, or use it for post-processing analysis to validate potential equilibrium solutions. 

Finally, divertor solutions optimized using field line diffusion models should be validated with higher fidelity codes, like EMC3-EIRENE \cite{feng_recent_2014}. EMC3-EIRENE has a more realistic plasma model, and takes into account neutral effects. However, these should not be used to replace field line diffusion models in the optimization loop due to their high computational cost.

\funding{This research was supported and funded by Type One Energy Group, Inc. © 2025 Type One Energy Group, Inc. All rights reserved. This research used resources of the National Energy Research Scientific Computing Center (NERSC), a Department of Energy Office of Science User Facility using NERSC award ERCAP0031926.}

\bibliographystyle{unsrtnat}
\bibliography{divertor}

\end{document}